# Subphases in the superconducting state of CeIrIn$_5$ revealed by low-temperature *c*-axis heat transport

H. Shakeripour[1,2,*] M. A. Tanatar[2,3,4,†] C. Petrovic,[5,6] and Louis Taillefer[2,6,‡]

[1]*Department of Physics, Isfahan University of Technology, Isfahan 84156-83111, Iran*
[2]*Département de physique and RQMP, Université de Sherbrooke, Sherbrooke, Canada J1K 2R1*
[3]*Ames Laboratory USDOE, Ames, Iowa 50011,USA*
[4]*Department of Physics and Astronomy, Iowa State University, Ames, Iowa 50011, USA*
[5]*Department of Physics, Brookhaven National Laboratory, Upton, New York 11973, USA*
[6]*Canadian Institute for Advanced Research, Toronto, Ontario M5G 1M1*

Low-temperature (down to ∼50 mK) thermal conductivity measurements with the heat flow direction along the interplane tetragonal *c* axis, $\kappa_c$, were used to study the superconducting state of heavy fermion CeIrIn$_5$. Measurements were performed in the magnetic fields both parallel to the heat flow direction $H \parallel c$, and transverse to it $H \parallel a$. Interplane heat conductivity in $H \parallel c$ configuration shows negligible initial increase with magnetic field and a rapid rise on approaching $H_{c2}$ from below, similar to the expectations for the superconducting gap without line nodes. This observation is in stark contrast to monotonic increase found in the previous in-plane heat transport measurements. In the configuration with the magnetic field breaking the tetragonal symmetry of the lattice $H \parallel a$, $\kappa_c$ reveals nonmonotonic evolution with temperature and magnetic field suggesting subphase boundary in the superconducting state. The characteristic temperature $T_{\text{kink}} \sim 0.07$ K of the subboundary is well within the domain of bulk superconductivity $T_c \sim 0.4$ K and $H_{c2} \sim 1.0$ T. These results are consistent with a superconducting gap with an equatorial line node and polar point nodes, a gap symmetry of the D$_{4h}$ point group, for which magnetic field along the tetragonal plane breaks the degeneracy of the multicomponent order parameter and induces a phase transition with nodal topology change.

## I. INTRODUCTION

Abundant magnetic fluctuations in the vicinity of a quantum critical point, at which the long-range magnetic order is suppressed by nonthermal tuning parameter [1], can serve as a glue for the superconducting pairing, leading to an unconventional superconductivity [2,3]. This mechanism is actively discussed as a scenario for superconductivity in the cuprates [4], where the quantum critical point is observed at the optimal doping with the highest $T_c$ [5,6]. Similarly, the existence of a magnetic quantum critical point under the superconducting dome close to the optimal doping was shown in the iron-based superconductors [7,8] and in the heavy fermion materials [9].

Contrary to conventional superconductivity, the magnetically mediated superconductivity can have a complex order parameter, characterized by the nodes in the gap structure and a possibility of degenerate ground states. Close proximity between the superconductivity and the magnetism suggest their relation, but does nearby magnetism directly affect the superconducting pairing? The difference in the characteristic $Q$ vectors of various magnetic structures, reflecting either Fermi surface topology (nesting) or local moment arrangements, can lead to the different patterns of hot spots on the Fermi surface and thus make different pairing states energetically favorable. The relation between the Fermi surface topology and the nesting $Q$ vector is believed to be responsible for $s\pm$ pairing in iron based superconductors [10], as opposed to $d$-wave pairing in the cuprates [11]. However, direct connection of the superconducting gap structure and the magnetic ordering pattern is lacking.

It is interesting that in the composition phase diagram of heavy fermion superconductors Ce(Rh$_{1-x}$Ir$_x$)In$_5$ the superconducting states of Rh and Ir compounds are disconnected [12], bordering different types of magnetic ordering. This observation suggests possible existence of two different superconducting orders. Although the superconducting state of CeRhIn$_5$ is believed to be $d$-wave, with the possibility of nodal $s$-wave order parameter [13], a controversy surrounds CeIrIn$_5$, with the suggestions of $d$-wave and multicomponent orders. In this article we report finding of the subphases in the superconducting state of CeIrIn$_5$, a key finding establishing the existence of two different types of pairings in the proximity to two different patterns of magnetic ordering.

The compounds with general formula Ce$M$In$_5$, where $M$ = Co, Rh, Ir, discovered in the early 2000s [14–16] and referred usually as 115 compounds, form fruitful playground for studying unconventional superconductivity. At ambient pressure CeRhIn$_5$ orders magnetically below $T_N = 3.8$ K and superconductivity is found deep in the magnetically ordered state below $T_c \sim 0.1$ K [17,18]. Superconductivity with $T_c \sim 2.5$ K is induced by pressure of the order of 2 GPa [16], not far from the pressure-tuned quantum critical point of magnetic order [1]. Heat capacity measurements in the pressure-induced superconducting state with the magnetic field rotating in the tetragonal (001) plane found fourfold anisotropy consistent with the $d$-wave pairing [19]. CeCoIn$_5$ is superconducting at ambient pressure with $T_c = 2.3$ K and the unconventional superconductivity mechanism (see Ref. [20] for review). This superconducting state is also proximate to magnetism, and the long-range magnetic order can be actually induced by Cd and Hg substitution of In [21], with a simple antiferromagnetic order $Q = (1/2, 1/2, 1/2)$ [22] different from spiral magnetic order of CeRhIn$_5$ $Q = (1/2, 1/2, 0.297)$ [23]. CeIrIn$_5$ shows bulk superconductivity below 0.4 K as found in heat capacity [24] and thermal conductivity [25,26] measurements. Long-range magnetic order in CeIrIn$_5$ is induced by Cd and Hg substitution similar to CeCoIn$_5$ [27], and has the same ordering $Q$ vectors as CeCoIn$_5$ [28]. The magnetism can be suppressed by pressure [29]. The calculated band structures [30] and measured Fermi surfaces [31,32] of CeCoIn$_5$ and CeIrIn$_5$ are similar. The main sheets of the Fermi surface are found to be nearly cylindrical, reflecting the unique tetragonal structure. Simultaneously small anisotropy of the electrical resistivity in both compounds [33] suggests the importance of the three-dimensionality of the Fermi surface. The observed branches are well explained by the $4f$-itinerant band model [31].

*hshakeri@cc.iut.ac.ir
†tanatar@ameslab.gov
‡Louis.Taillefer@USherbrooke.ca

Both the anisotropy of thermal conductivity finding finite residual linear term for the in-plane transport and zero residual term for the interplane transport [25] and the universal response of the in-plane versus increasing response to the natural disorder [34] in CeIrIn$_5$ suggest superconducting gap with equatorial line node and polar point nodes. This gap structure is consistent with $E_g(1, i)$ state having $(x + iy)z$ basic function. This state is different from the $d_{x^2-y^2}$ state as suggested by the fourfold heat capacity oscillations in CeRhIn$_5$. However, our conclusion about the structure of the superconducting gap in CeIrIn$_5$ was argued against based on the results of thermal conductivity measurements down to the temperatures of about 100 mK [26] and the heat capacity measurements down to 80 mK [35] in magnetic fields rotating in the tetragonal plane, interpreted as an evidence for the $d$-wave state with vertical line nodes. A similar fourfold variation of the heat capacity on magnetic field rotation was found in the pressure-induced superconducting state of CeIrIn$_5$ [36]. London penetration depth measurements find $T-$linear dependence down to 80 mK [37], consistent with the superconductor with line nodes. It was argued that the $c$-axis thermal conductivity can be explained in the $d$-wave model for special two-dimensional Fermi surface [38].

Additional feature of CeIrIn$_5$ is the observation of zero-resistance state below approximately 1 K, well above bulk $T_c = 0.4$ K. Extensive studies of anisotropic resistivity in microstructures fabricated using focused ion beam cutting [39] found that zero-resistance state is observed only in the interplane resistivity measurements, but not in-plane resistivity measurements.

Motivated by these interesting observations of the directional character of the zero-resistance state and the discrepancy with higher temperature in-plane heat transport measurements, here we revisit study of the superconducting state in CeIrIn$_5$ using thermal transport with heat flow along the tetragonal $c$ axis as a directional tool to study the superconducting gap. We found nonmonotonic field and temperature dependence of $\kappa_c$ in magnetic field parallel to conducting plane suggesting existence of subphases in the superconducting state of CeIrIn$_5$. Experimentally found rapid increase of the thermal conductivity in the magnetic field $H \parallel a$ is consistent with the field-induced formation of vertical line nodes as expected for multicomponent order parameter. For the temperatures above bulk $T_c = 0.4$ K but below zero-resistance $T_{c,\rho} = 0.8$ K we find $\kappa_c$ to be independent of the field value in longitudinal $H \parallel c$ configuration. We find that thermal conductivity in $T \to 0$ limit is field independent above bulk $H_{c2} = 0.5$ T and at field $H \parallel c = 4$ T, above resistive $H_{c2\rho}$ obeys the Wiedemann-Franz law. Both these observations are inconsistent with the expectations for bulk superconductors.

## II. EXPERIMENTAL

Single crystals of CeIrIn$_5$ were grown by the self-flux method [15]. The sample studied in great detail in this study was the sample with the lowest residual resistivity in our disorder-dependent $c$-axis heat transport study [34]. The main focus in this paper is put on the sample with the lowest $\rho_{0c} = 0.5$ $\mu\Omega$ cm as found in $T \to 0$ and $H \to 0$ limit (see Fig. 2 below). The features were still observed in the second best sample, but were somewhat smeared. The bulk transition temperature for pure samples is $T_c = 0.38 \pm 0.02$ K and the upper critical field $H_{c2} = 0.49$ T and 1 T for $H \parallel c$ and $H \perp c$, respectively [35].

The samples were mechanically cut and polished to have the shape of parallelepipeds with dimensions $\sim 1 \times 0.15 \times 0.086$ mm$^3$ and $c$ axis as the longest direction. Contacts to the samples were made with Ag wires soldered with indium/silver alloy. Same contacts were used in both electrical resistivity $\rho$ and thermal conductivity $\kappa$ measurements. This way we essentially eliminated big uncertainty of geometric factor determination in quantitative comparison of the two quantities [40] to verify Wiedemann-Franz law, $\kappa/T = L_0/\rho$, where $L_0 = 2.45 \times 10^{-8}$ [W$\Omega$/K$^2$] is Sommerfeld value of Lorenz number. These contacts have typical low-temperature resistance of $\sim 1$ m$\Omega$. Low resistance of the contacts is crucial for proper measurements of thermal conductivity at the lowest temperatures [41] due to potential electron phonon-decoupling. Verification of the Wiedemann-Franz law in the normal state in our samples (see Fig. 2 below) provides the best test for correct thermal conductivity measurements. Samples were cooled in magnetic field from above $T_c$ to ascertain homogeneous field distribution in the superconducting state.

The thermal conductivity was measured in a dilution refrigerator using a standard four-probe one-heater two-thermometers steady-state method with two RuO$_2$ chip thermometers calibrated *in situ* against a reference RuO$_2$ thermometer [42]. For the following discussion it is important to note that all thermal conductivity data presented in this paper represent electronic contribution. Phonon conductivity in the interplane heat flow configuration was measured in our previous study [34], following protocol discussed in Ref. [43]. It represents 3% correction at 1 K, and is notably below this below 0.4 K, the main focus of this study. For reference we show phonon contribution $\kappa_{ph}$ in Fig. 1(a).

## III. RESULTS

### A. Magnetic field parallel to the tetragonal $c$ axis $H \parallel c$

In Fig. 1 we show temperature-dependent interplane thermal conductivity of CeIrIn$_5$, measured in the magnetic fields parallel to the tetragonal $c$ axis $H \parallel c$. Blue curve in panel (a) shows for reference the phonon contribution to thermal conductivity $\kappa_{ph}/T$. The effects of magnetic field on thermal conductivity in the normal state are minimum in this configuration, due to the lower values of the upper critical fields and the longitudinal alignment of field and heat current. In the top panels the thermal conductivity is presented as $\kappa_c/T$ versus $T$ over the whole range up to zero-resistance $T_c = 0.8$ K [Fig. 1(a)] and zoom over range of bulk superconductivity $T_c = 0.4$ K [Fig. 1(b)], indicated with arrow. In Fig. 1(c) the data are plotted normalized by the normal state 0.5 T curve, $\kappa/\kappa_N$, clearly showing branching of the curves below bulk $T_c = 0.4$ K. Figure 1(d) shows the field-dependent normalized thermal conductivity $\kappa_s/\kappa_N$ taken at 0.1 K (solid-olive up-triangles), at base temperature of our experiment 56 mK (black-open circles) and in $T \to 0$ extrapolation (blue-solid circles).

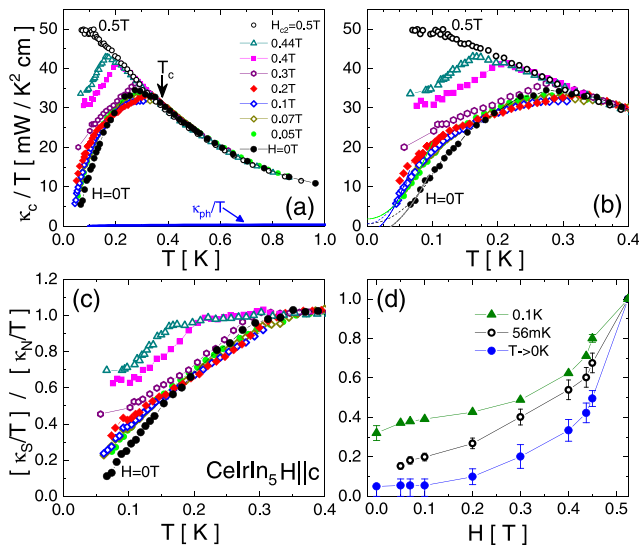

FIG. 1. Temperature-dependent interplane thermal conductivity, plotted as $\kappa_c/T$ vs $T$, over the whole range up to zero-resistance $T_c = 0.8$ K (a) and over range of bulk superconductivity $T_c = 0.4$ K (b) in magnetic fields parallel to $c$ axis along the heat current. Blue line in panel (a) shows phonon contribution $\kappa_{ph}/T$. Bottom to top: 0 T (black-solid circles), 0.05 T (green-solid circles), 0.07 T (dark-yellow open diamonds), 0.1 T (blue-open diamonds), 0.2 T (red-solid diamonds), 0.3 T (magenta-open hexagons), 0.4 T (magenta-solid squares), 0.44 T (olive-open up-triangles), and 0.5 T (normal state, black-open circles). In (c) the data are plotted normalized by the normal state 0.5 T curve, $\kappa/\kappa_N$, clearly showing convergence of curves above bulk $T_c = 0.4$ K. (d) Field-dependent normalized thermal conductivity $\kappa_s/\kappa_N$ taken at 0.1 K (olive-solid up-triangles), at base temperature of our experiment 56 mK (black-open circles) and in the $T^2$ extrapolation $T \to 0$ (blue-solid circles).

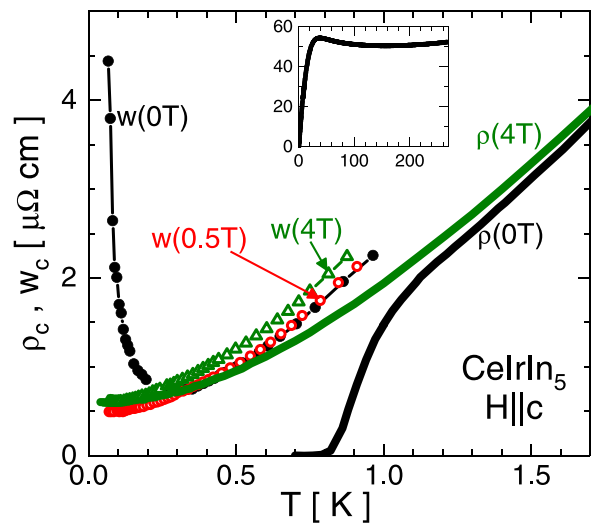

FIG. 2. Temperature-dependent interplane resistivity, measured in zero magnetic field (black-solid line) and in magnetic field 4 T parallel to $c$ axis (above resistive $H_{c2\rho}$, olive-solid line). For comparison we plot thermal analog of the electrical resistivity, $w \equiv L_0 T/\kappa$ in zero field (black-solid circles) and in the magnetic fields above bulk $H_{c2} = 0.5$ T (red-open circles) and resistive $H_{c2\rho} = 4$ T (olive-open up-triangles). The $w$ and $\rho$ curves for $H = 4$ T converge at $T \to 0$ as expected for the Wiedemann-Franz law. The $w(T)$ data at 0.5 T show the same temperature dependence as 4 T curve shifted to the lower values due to the magnetoresistance. Inset shows $\rho(T)$ curve over the whole temperature range up to room temperature.

Several features of the data of Fig. 1 need to be mentioned. While crossing the bulk $T_c$ does not produce any visible feature for the zero-field curve, measurements in finite magnetic fields show a clear feature at $T_c$ [best seen in panel (c) removing background temperature dependence of the normal state]. This feature enables determination of the bulk $H_{c2}(T)$. The difference between the smooth zero-field curve and the in-field measurements comes from scattering of quasiparticles on vortices, a behavior seen in both conventional [44,45] and unconventional [46–48] superconductors.

In the magnetic field of 0.5 T (above bulk $H_{c2}$), $\kappa_c/T$ shows monotonic increase on cooling, as expected for a metal with significant contribution of the inelastic scattering even at these low temperatures, similar to in-plane transport [49]. The data above bulk $T_c = 0.4$ K stay on top of each other for all field values, as expected for a normal metal with negligible magnetoresistance but not a bulk superconductor. In Fig. 2 we make direct comparison of thermal conductivity measurements for various characteristic fields with resistivity measurements in zero magnetic field and in the magnetic field of 4 T above the resistive $H_{c2\rho}$. The thermal conductivity data are presented as thermal analog of electrical resistivity $w \equiv L_0 T/\kappa$. Presenting data this way makes it visually easy to check the validity of the Wiedemann-Franz law, which is satisfied when $w = \rho$. As can be seen from Fig. 2, the $w(4T)$ and $\rho(4T)$ curves converge in $T \to 0$ limit, clearly showing that the WF law is obeyed. Interestingly, the shift between $w(0.5T)$ and $w(4T)$ curves in $T \to 0$ limit is about the same as shift of resistivity between $\rho(0T)$ and $\rho(4T)$ curves above $T_c$ clearly showing its origin in the normal state magnetoresistance. This observation reveals negligible contribution of the superconducting carriers to the heat transport in $T \to 0$ limit in a range between $H_{c2}$ and $H_{c2\rho}$. Both observations of the negligible magnetic field effect on heat transport between bulk $T_c$ and $T_{c\rho}$ in zero field, Fig. 1, and between bulk $H_{c2} = 0.5$ T and $H_{c2\rho} = 4$ T at low temperatures, Fig. 2, are suggesting nonbulk character of superconductivity in the 1 K superconducting phase observed by resistivity measurement (Fig. 2).

In Fig. 1(d), we plot the field-dependent thermal conductivity normalized to the normal state value $\kappa_n = \kappa_c(0.5 \text{ T})$. We show the actual data at 0.1 K (green up-triangles) and 56 mK (the base temperature of our measurements, open-black circles). We also plot data extrapolated to $T \to 0$ using a procedure described in our previous studies [25,34]. As can be seen in Fig. 1(b), the linear extrapolation of $\kappa_c/T = \kappa_0/T + \kappa_1/T * T$ gives physically meaningless negative value of $\kappa_0/T$. This simple fact suggests that the higher power terms, $\kappa_c/T = \kappa_0/T + \kappa_2/T * T^2$, should be dominant at low temperatures. This is different from $T$−linear variation of $\kappa/T$ on approaching $T = 0$ limit for in-plane transport [25] in line with theoretical expectation [50]. The $T^2$ term is expected in the superconductors with point nodes [50,51] and is indeed observed in samples with higher residual resistivity [34]. Using simple quadratic form $\kappa_c/T = \kappa_0/T + \kappa_2/T * T^2$ we should get the the highest estimate of the residual linear term $\kappa_0/T$. We were applying this procedure to fit a few lowest temperature data points, and used the average as the experimentally extrapolated value. Using this procedure we obtain residual term in the range 1–2 mW/cmK$^2$ (see Fig. 5 below) for all fields below 0.1 T and rapid increase of $\kappa/\kappa_N$ on approaching $H_{c2}$. This is in stark contrast with the monotonic increase found in the in-plane heat transport from finite value in zero field [25,34]. The trend for a flat-field dependence can be tracked in the actual data taken at base temperature of our experiment, 56 mK.

Note an extended range up to magnetic field 0.2 T, in which $T \to 0$ extrapolated $\kappa_s/\kappa_N$ value remains zero within error bars of extrapolation; see Fig. 1(c).

### B. Magnetic field parallel to the tetragonal plane *a* axis $H \parallel a$

In Fig. 3 we show temperature-dependent interplane thermal conductivity of CeIrIn$_5$, measured in the magnetic fields perpendicular to the tetragonal *c* axis along the *a* axis in the conducting plane $H \parallel a$. Higher value of the bulk upper critical field in this configuration, $H_{c2a} = 1$ T [35], makes it necessary to use higher magnetic fields, which cause a notable magnetoresistance in this transverse to the heat current configuration. In Fig. 3(a) we present thermal conductivity $\kappa_c/T$ versus $T$ over the whole range up to the zero-resistance $T_c = 0.8$ K. It can be seen that the data above bulk $T_c = 0.4$ K clearly show systematic down shift due to contribution of the normal state magnetoresistance. This down shift makes it hard to precisely determine $H_{c2}$ from the thermal conductivity measurements themselves from the point of deviation from the normal state curve, as done in $H \parallel c$ configuration in Fig. 1. The lowest estimate of the $T_c(H)$ can be obtained from the positions of the maximums, below which $\kappa_c/T$ shows rapid decrease. Of note the shape of the curve in $H = 1$ T. It shows nonmonotonic increase of $\kappa/T$ with the rapid rise below 0.1 K. This rise presumably reflects the combined effect of the normal state magnetoresistance (progressively pushing all $\kappa/T$ curves above $T_c$ down) and the leftover of $\kappa/T$ rise below $T_c$, which is still not suppressed to zero at this field. Figure 3(b) zooms the low-temperature portion of the $\kappa_c/T$ vs $T$ curves. With application of the small magnetic field above 0.05 T the curves flatten and reveal obviously nonmonotonic temperature dependence with pronounced downturn starting at a characteristic temperature $T_{\text{kink}}$ approximately 70 mK [arrows in Fig. 3(b)], which is suppressed by the magnetic field. This sharp feature can signal a phase change in the superconductor with multicomponent order parameter, and delineates a new domain inside the superconducting phase, as summarized in the $H - T$ phase diagram below, Fig. 7. Normalization of the $\kappa_c/T$ data by the 1 T field curve in panel (c) leads to a 20% increase above the normal state value in small fields due to the effect of magnetoresistance and introduces feature at 0.1 K in $H = 0.8$ T curve, very similar to the kink features for the lower fields. This observation may be suggestive that the kink line extends all the way to $H_{c2}$ line.

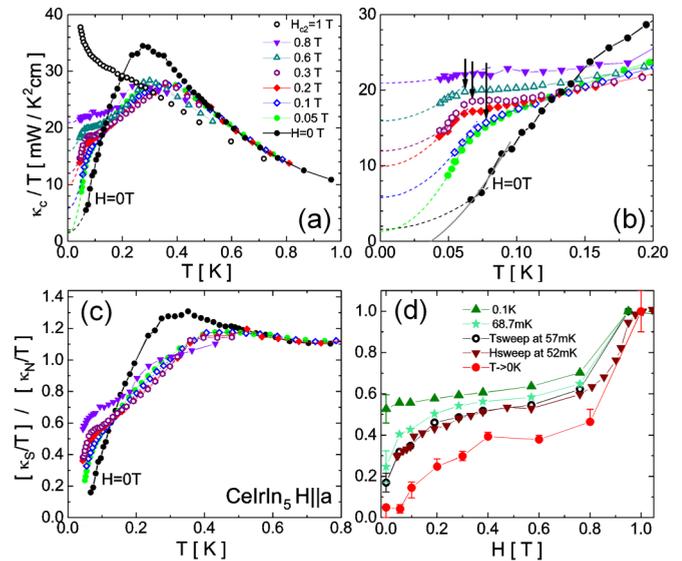

FIG. 3. Temperature-dependent interplane thermal conductivity, plotted as $\kappa_c/T$ vs $T$, over the whole temperature range up to zero-resistance $T_c = 0.8$ K (a) and zoom at the lowest temperatures below 0.2 K (b) in magnetic fields parallel to the *a* axis in the tetragonal plane transverse to the heat current. Bottom to top: 0 T (black-solid circles), 0.05 T (green-solid circles), 0.1 T (blue-open diamonds), 0.2 T (red-solid diamonds), 0.3 T (magenta-open hexagons), 0.6 T (olive-open up-triangles), 0.8 T (purple-solid down-triangles), and 1 T (normal state, open-black circles). Dashes show $T^2$ fit of the last two data points for all field values, grey-solid line shows fitting range dependence of the $T \to 0$ extrapolation using last four data points for zero-field curve, illustrating the error bar evaluation. Arrows in panel (b) show onset of the rapid decrease of thermal conductivity below the characteristic temperature $T_{\text{kink}}$. In the left-bottom panel (c) the data are plotted normalized by the normal state curve at 1 T, $\kappa/\kappa_N$. Bottom-right panel (d) shows field-dependent normalized thermal conductivity $\kappa_s/\kappa_N$ taken at 0.1 K (olive-solid up-triangles), at 68 mK (light-cyan stars), at base temperature of our experiment 57 mK in $T$− sweep mode (black-open circles), in $H$− sweep mode at 52 mK (brown-solid down-triangles), and in the $T^2$ extrapolation to $T \to 0$ (red-solid circles).

In the Fig. 3(d) we summarize evolution of the field-dependent thermal conductivity with temperature in $H \parallel a$ configuration. We use the normalized thermal conductivity scale $\kappa/\kappa_N$ and the normalized magnetic field scale $H/H_{c2}$. We show the data for 0.1 K (above the $T_{kink}$, solid-olive up-triangles), $T = 68$ mK (slightly below $T_{kink}$, cyan stars), $T = 57$ mK (base temperature in $T$-sweep measurements, black-open circles), $T = 52$ mK (measurements in $H$-sweep mode, brown down-triangles) and in $T \to 0$ extrapolation (red-solid circles). Because of the very steep decrease of thermal conductivity at the lowest temperatures, the extrapolation to $T \to 0$ were made assuming a $T^2$ dependence of $\kappa_c/T$ ($T$-linear extrapolations provide physically meaningless negative values at low fields). The $\kappa_c/T \propto T^2$ comes from the contributions of the point nodes at the poles of the Fermi surface (see Fig. 7 below). We fitted a few low $T$ data points (2 to 5 points) with this function, as shown by the dashed lines in Fig. 3(b). The error bars are determined by the minimum and maximum values of extrapolated residual linear terms as shown for zero-field curve with the grey line). For temperatures below $T_{kink}$ thermal conductivity rises notably faster at the lowest fields and a range of zero residual linear term in thermal conductivity in $T \to 0$ extrapolation terminates at very low field of 0.1 T.

### C. Comparison of the two magnetic field directions

Comparison of Fig. 1(d) and Fig. 3(d) reveals that the thermal conductivity rises notably faster in the $H \parallel a$ than in $H \parallel c$ configuration. An explicit comparison is made in Fig. 4, using normalised $\kappa/\kappa_N$ vs $H/H_{c2}$ plots. To make sure that the difference in response in the two experimental configurations is not an artefact of the extrapolation procedure, we plot data at base temperatures of the experiments (open symbols) and in the $T^2$ extrapolations to $T = 0$ (solid symbols). The anisotropy between the data at 56 mK in $H \parallel c$ (open-blue squares) and at 57 mK in $H \parallel a$ (open-red circles) is strongly magnetic field dependent and at the maximum near $H/H_{c2} = 0.2$ it is bigger that the factor of 2. This anisotropy ratio should be contrasted with tiny few percent variations in the response to the rotation of the magnetic fields in the plane [26]. In $T \to 0$ extrapolations the difference becomes even bigger, and it reveals notably different field dependence for $H \parallel c$ and $H \parallel a$.

To get an additional insight into the origin of the difference, in Fig. 5 we explicitly compare $\kappa_c/T$ versus $T$ in the magnetic fields corresponding to a maximum anisotropy $H/H_{c2} = 0.2$. This figure reveals a clear correlation between the appearance of the kink in the $T$-dependence and the rapid rise in the $H$ dependence. Of note, although $T$-linear extrapolation of the $\kappa_c/T$ versus $T$ to $T \to 0$ in $H \parallel a$ gives smaller residual linear term than $T^2$ extrapolation, the actual data points in the $T$-sweep measurements in $H \parallel c$ configuration, stay below the $T$-linear extrapolation, clearly ruling out that the appearance of the anisotropy is an artifact of the extrapolation procedure used.

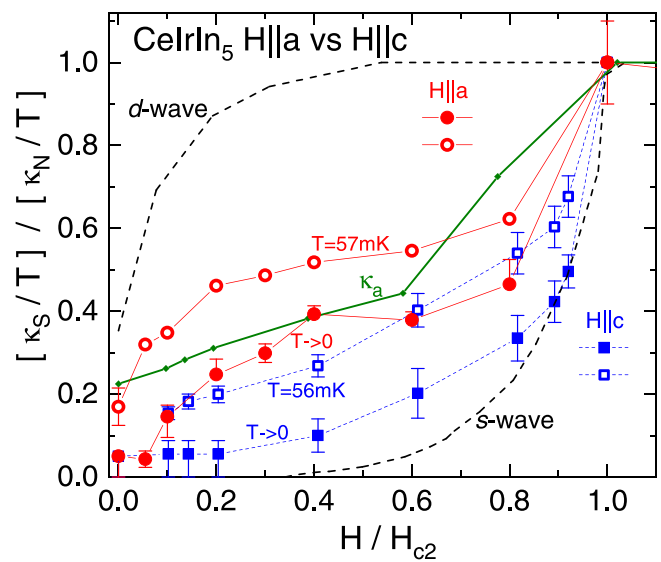

FIG. 4. Residual linear terms determined with $T^2$ extrapolation of the data (solid symbols) $\kappa_0/T$ in the $c$-axis thermal conductivity of CeIrIn$_5$, plotted on scales normalized to the normal state $\kappa/\kappa_N$ vs $H/H_{c2}$. Red circles are for $H \parallel a$, blue squares for $H \parallel c$. Green-solid line shows the field dependence for the in-plane heat transport in $H \parallel c$ configuration [52]. For reference we plot the actual data at the base temperatures of our experiment (open symbols) and the standard dependencies (dashes) observed in the $s$-wave superconductors, as in the clean Nb and the dirty InBi shown here (reproduced from Ref. [52]) and for the $d$-wave (nodal) superconductor Tl-2201 [53].

In Fig. 4 we plot for reference with dashes the standard dependencies observed in isotropic $s$-wave superconductors, as in the clean Nb and the dirty InBi shown here (reproduced from Ref. [52]) and for the $d$-wave (nodal) superconductor Tl-2201 [53]. The $T = 0$ extrapolation curve for $H \parallel c$ is close to an activated dependence in the full-gap *s*-wave superconductors. It is notably flatter than the dependence in the in-plane heat transport (solid-green line [52]), which shows a finite residual linear term in zero field [25] and rises monotonically with the field as expected for superconductors with the line nodes. Although the precise shape of the curve in $T = 0$ extrapolation can depend on the extrapolation procedure in $H \parallel a$ configuration, as we discussed above for $H/H_{c2} = 0.2$ curve, the rapid rise in thermal conductivity above $H/H_{c2} = 0.05$ is real and is much faster than in the $H \parallel c$ configuration. This nonmonotonic increase with the magnetic field in $H \parallel a$ suggests development of the vertical lines nodes in the magnetic fields, reflecting transformation of the superconducting gap nodal structure, as expected for the hybrid superconducting gap.

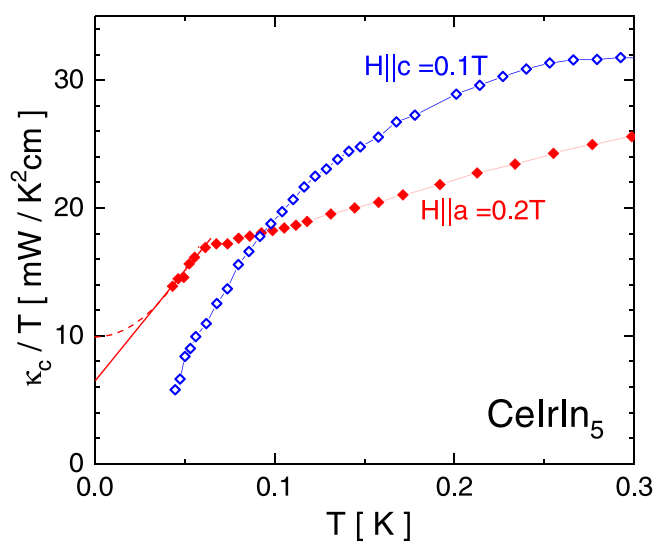

FIG. 5. Comparison of the $c$-axis thermal conductivity $\kappa_c/T$ of CeIrIn$_5$ taken in equivalent normalized magnetic fields $H/H_{c2} \sim 0.2$ for which the difference between the two configurations is maximum, Fig. 4, in $H \parallel c = 0.1$ T and $H \parallel a = 0.2$ T. The difference in the behavior for the two field configurations is obvious.

The difference in the field dependence between the in-plane and the interplane thermal conductivity in $H \parallel c$ agrees with the superconducting gap with horizontal line node. The interplane conductivity shows zero (within error bars) residual linear term in zero magnetic field and a significantly slower increase with the increase of magnetic field. The residual term in zero magnetic field in the in-plane thermal conductivity does not depend on the residual resistivity of the samples (universal thermal conductivity) [34] as expected for the superconductors with line nodes. The interplane thermal conductivity strongly increases with the residual disorder, as expected for the superconductors without line nodes. Both these differences support hybrid superconducting gap structure with the polar point nodes and a horizontal line node at the lowest temperatures.

## IV. DISCUSSION

### A. Comparison with heat capacity measurements

To a notable extent our study was motivated by the discrepancy of our observation of zero residual linear term in *c*-axis heat transport measurements with the results of the angular-dependent thermal conductivity [26] and heat capacity [35] measurements. While the thermal conductivity study was conducted at temperatures significantly higher than $T_{kink}$, the base temperature in heat capacity measurements (80 mK) is close to $T_{kink}$. In Fig. 6 we directly compare the field-dependent interplane thermal conductivity with the heat capacity data of Kittaka *et al.* [35]. It is interesting that the onset of the rapid rise in thermal conductivity roughly coincides with the slope-change feature in heat capacity measurements. This feature was fitted by Kittaka *et al.* with square-root dependence expected for the nodal superconductors [54,55]. Note that because of the high base temperature $0.2T_c$, zero-field value in heat capacity measurements amounts to nearly half of the normal state value and the feature was most likely overlooked in the data.

### B. Comparison with CeCoIn$_5$

It is of interest to put together finding of this and our previous [25,34] thermal conductivity measurements in CeIrIn$_5$ with the studies in the closely related CeCoIn$_5$. Thermal conductivity of CeCoIn$_5$ reveals very pronounced multiband effects [56,57]. While at temperatures as low as 50 mK $\kappa/T$ extrapolates to sizable residual linear term, strongly suppressed with disorder [56], this term diminishes to a very small value on further cooling [57]. This term was assigned to a small part of the Fermi surface, accounting for approximately 15% of the density of states, with extremely small superconducting gap. Application of a magnetic field parallel to the plane rapidly suppresses the residual term in $\kappa/T$ in small field [58]. It was suggested recently that this group of carriers may be due to the Dirac fermion part of the Fermi surface [59].

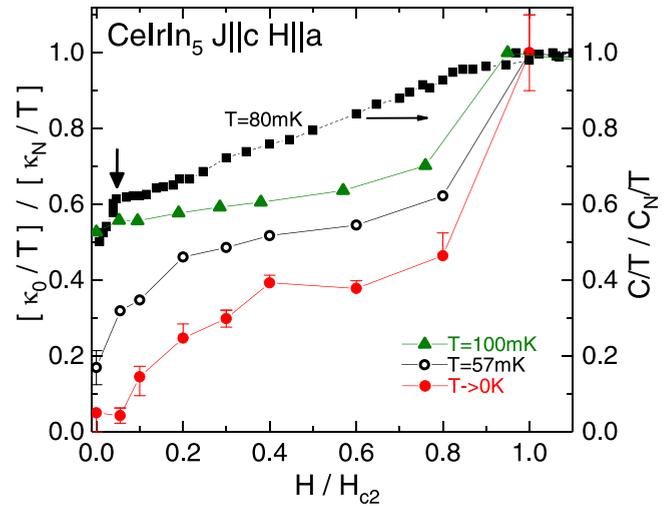

FIG. 6. Comparison of the field-dependent interplane thermal conductivity $\kappa_c$ with the heat capacity measurements of Ref. [35] both taken in the configuration with magnetic field parallel to the plane $H \parallel a$. For both measurements we normalize the data by their normal state values and use a dimensionless field scale $H/H_{c2}$. Relatively high base temperature in the heat capacity measurements, 80 mK or 0.2 $T_c$, leads to the value of the residual heat capacity in zero field amounting to the half of the normal state value. For reference we show thermal conductivity taken at the closest temperatures 0.1 K (solid triangles) and 57 mK. Note sharp feature in the field-dependent heat capacity at small fields, denoted with arrow, and plotted in the phase diagram in Fig. 7 below.

The response of CeIrIn$_5$ to all types of the perturbations is dramatically different from CeCoIn$_5$. Disorder increases residual linear term for the interplane transport [34], while the in-plane transport shows a universal doping independent value. Application of the magnetic field rapidly increases the residual linear term in the in-plane transport, as opposed to it suppression in CeCoIn$_5$. Both these results are in line with expectations of the nodal superconductors, and do not indicate the presence of uncondensed carriers.

### C. Phase diagram

#### 1. Theory considerations

The allowed superconducting order parameter representations in the tetragonal symmetry for a singlet pairing [11,25] have 6 nodal states in addition to the nodeless $A_{1g}$ *s*-wave state. The nodal $A_{2g}$ *g*-wave $xy(x^2 − y^2)$ state and two *d*-wave states, $B_{1g}$ $x^2 − y^2$ and $B_{2g}$ $xy$, have vertical line nodes. Hybrid $E_g(1, 0)$ $xz$ and $E_g(1, 1)$ $(x + y)z$ states have vertical and horizontal line nodes, hybrid $E_g(1, i)$ $(x + iy)z$ state has horizontal line nodes at the equator and point nodes at the poles. This state is the most consistent with our low-temperature phase in zero field and is also one of the most promising candidates for the superconducting gap of UPt$_3$ [51].

### 2. Superconductivity with multicomponent order parameter

According to theory, the degeneracy contained in the two-component order parameter of a superconductor can be lifted by applying a magnetic field away from the high-symmetry $c$ axis of the tetragonal crystal structure. A number of different Ginzburg-Landau models have been proposed, most of which require an unconventional superconducting order parameter. Two most plausible models, are based on: (i) a single multi-component order parameter coupled to a symmetry breaking field [60–62]. Here, the degeneracy of a two-dimensional even or odd parity order parameters is lifted by a symmetry breaking field, and (ii) theories based on two symmetry unrelated order parameters, which are accidentally nearly degenerate [63–65].

To explain more, in the former model, so called 2D $E$-representation model, the theory has only one phase transition in zero field and by itself cannot explain the double transition [60]. The splitting is caused by lifting of the degeneracy of a two-component superconducting order parameter by a symmetry-breaking field. In this model, the possible 2D representations of even-parity (spin-singlet) pair states or odd-parity (spin triplet) symmetry in a tetragonal crystal structure with point group D$_{4h}$ are $E_g$ or $E_u$, respectively. A superconducting order parameter belonging to one of these representations can be represented by a complex vector $\vec{\eta} = (\eta_1, \eta_2)$, whose components are the coefficients multiplying the basis functions $\psi$ of the two dimensional representation:

$$\Delta(\mathbf{k}) = \eta_1 \psi_1(\mathbf{k}) + \eta_2 \psi_2(\mathbf{k}). \quad (1)$$

Despite all $E$ representation models are based on two-component orbital order parameters, it was reported that they yield different predictions for the thermodynamic, magnetic and transport properties, including the $H − T$ phase diagram [61].

Symmetry arguments imply that the vortex lattice phase diagram contains at least two vortex lattice phases for magnetic fields applied along any of the symmetry axes in the $ab$ plane: (1,0,0), (0,1,0),(1,1,0),(1,−1,0). To illustrate the origin of these phase transitions, consider a zero-field ground state $\vec{\eta} = (1, i)$ and a magnetic field applied along the (1,0,0) direction. Due to the broken tetragonal symmetry, the degeneracy of the $\vec{\eta} = (1, 0)$ and the $\vec{\eta} = (0, 1)$ solutions is removed by the magnetic field. Consequently only one of these two possibilities will order at the upper critical field [66]. When field is applied along $c$ axis, calculation predicts that $\vec{\eta} = (1, i)$ is stable (since this phase minimize the number of nodes in the order parameter) and no change in the symmetry [67,68].

In the latter model, so called the accidentally degenerate models, the phase diagram is accounted for by two primary order parameters belonging to different irreducible representations [65]. The splitting of the phase transition is due to accidental degeneracy, not to coupling to the magnetism. Once two representations are involved, the possibilities for the form of the order parameter become numerous. In the accidentally degenerated models the two representations can have the same or different parity [63].

In brief, for multicomponent order parameter superconductors external magnetic field acts similar to the small internal magnetic field. If the direction of magnetic field is different from the highest symmetry axis, the response of the superconductor is determined by lifting the degeneracy of the two order parameters, and a phase transition with nodal topology change is expected. Lifting the degeneracy creates vertical line nodes, so this topology change should be the most obvious for the heat current along $c$ axis. On the contrary, for magnetic field parallel to the high symmetry direction ($c$ axis), the response is the same as for single component order parameter superconductors, i.e., it should reveal nodal behavior for in-plane transport and activated behavior for interplane transport.

### 3. Summary of the experimental findings

In Fig. 7 we summarize the observations in the $H − T$ phase diagram of the superconducting state of CeIrIn$_5$ in $H \parallel a$ configuration. The low temperature-low magnetic field phase in the phase diagram shows the properties (strong $ac$ plane thermal conductivity anisotropy [25], universal versus nonuniversal response to disorder [34], difference in magnetic field response, see Fig. 4(a), consistent with $E_g(1, i)$ state (if a tetragonal crystal structure with spin-singlet even parity is considered), as discussed previously [25]. Its high-temperature boundary seems to correlate with the extension of $T_{\text{kink}}$ line to zero field, as the $ac$-plane anisotropy increase as a function of temperature [25] starts below 0.3 $T_c \sim$ 120 mK [25].

The state below the $T_{\text{kink}}$ line sets in magnetic fields higher than 0.05 T and has a notably increased thermal conductivity along $c$ axis. It is natural to relate it with the magnetic field induced transformation from the state with the horizontal line node $E_g(1, i)$ to a state with the horizontal and the vertical line nodes $E_g(1, 0)$ or $E_g(0, 1)$.

This phase diagram suggests that high-temperature state with the negligible $ac$-plane anisotropy of thermal conductivity in zero field [25] above 120 mK and the negligible anisotropy on magnetic field rotation in the $ac$ plane may indeed be a $d$-wave state, as suggested by the fourfold oscillations on the magnetic field rotation in the tetragonal plane.

The high-temperature phase, which shows a four fold symmetry in the tetragonal plane in the field-angle thermal conductivity [26] and heat capacity [35] measurements was suggested to have $B_{1g}$ $d$-wave symmetry [26]. It does not reveal characteristic anisotropy of the horizontal line node in the field-angle [26] and directional [25] thermal conductivity measurements and thus may have $E_g(1,0)$ or $E_g(1,1)$ symmetry or $d_{x^2−y^2}$ symmetry, if the accidental degenerate models are considered.

To interpret the whole phase diagram in Fig. 7 the accidentally degenerate models may be necessary; the system shows a $d$-wave symmetry (B$_{1g}$) at high temperatures and the hybrid gap symmetry [E$_g(1, i)$] at low temperatures. Theoretical work on the gap symmetry of CeIrIn$_5$ by Maki *et al.* [69] in the chiral $d$-wave state based on a weak-coupling BCS theory, at low temperatures, below $T \ll 0.3T_c \sim 0.12$K, shows a good

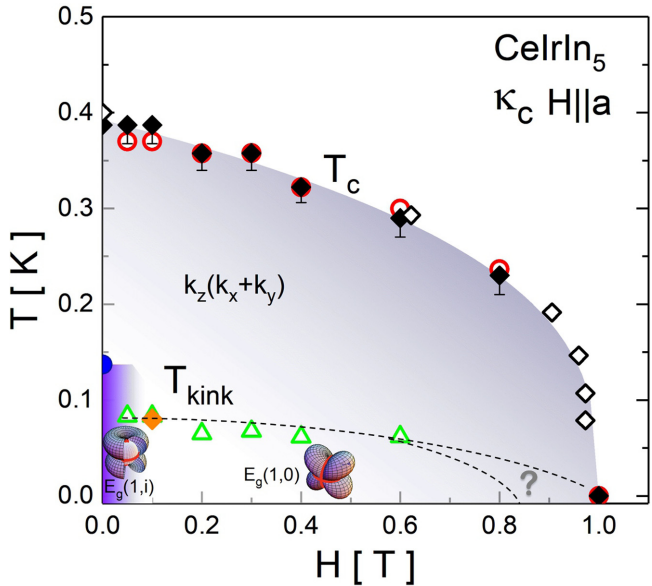

FIG. 7. The $H-T$ phase diagram of the superconducting state of CeIrIn$_5$ as revealed by the interplane heat transport in the magnetic fields parallel to the tetragonal conducting plane $H \parallel a$. The $T_c(H)$ points (red-open circles and solid diamonds) were determined using steep kinks at the transition temperature in $\kappa/T$, Fig. 3(a), or in $\kappa_s/\kappa_N$, Fig. 3(c), these data are in reasonable agreement with the heat capacity data (black-open diamonds) revealing much clearer anomalies at $T_c$ [35]. The temperature of the kink feature $T_{kink}$ (open up-triangles) was determined at onset of the decrease in $\kappa_c/T$ vs $T$ [arrows in Fig. 3(b)]. This feature is linked with the rapid rise of $T \to 0$ extrapolation in the field dependence. The low field-low temperature phase, below 0.07 K and 0.05 T, has strong $ac$ plane gap anisotropy [25,34], big blue point corresponds to an onset of the anisotropy increase on cooling. This suggests its relation to the hybrid $E_g(1, i)$ symmetry [25]. The orange-solid diamond point is from heat capacity data Ref. [35], see Fig. 6. The high field-low temperature phase has a notably increased residual conductivity, suggestive of transformation of $E_g(1, i)$ state to a lower symmetry $E_g(1, 0)$ or $E_g(0, 1)$ state with vertical line nodes. Whether the high field boundary of the phase (dashed lines) is going all the way to $H_{c2}$, or terminates before it, is not defined in our experiment. The high-temperature phase with fourfold symmetry in the thermal conductivity and the heat capacity measurements might have $d$-wave symmetry [26,35]. Note, the phase diagram is consistent with the theoretical accidental degeneracy models [63]. Although, according to the theoretical 2D representation models [60] $E_g(1, 0)$ or $E_g(1, 1)$ symmetries are also possible.

agreement with our observations and in contrast to the $d$-wave symmetry (see [69]).

### D. Subphases in other superconductors

There are few examples of the superconductors with multiple phases. High-field phases were observed below $H_{c2}$ lines in CeCoIn$_5$ [70,71] and Sr$_2$RuO$_4$ [47,72,73] only in magnetic field parallels to the conducting plane, similar to our observations in CeIrIn$_5$. Several low-field phases were observed in UPt$_3$ [46,51,74] and PrOs$_4$Sb$_{12}$ [75]. The interesting point is that UPt$_3$ and PrOs$_4$Sb$_{12}$ superconductors have point nodes in their gap symmetries, similar to CeIrIn$_5$.

The thermal transport and field-angle-dependent specific heat measurements of the heavy fermion compound URu$_2$Si$_2$, with the body centered tetragonal structure, shows hybrid $E_g$ gap symmetry [76,77]. The superconducting state of this compound below $T_c = 1.5$ K is observed within the hidden order phase stabilized below $T_N = 17.5$ K. The field dependence of the in-plane thermal conductivity of URu$_2$Si$_2$ is very similar to that observed in CeIrIn$_5$ [Fig. 4(a)], which was related to the response of the superconducting state with the line and point nodes (hybrid symmetry) under applied magnetic field [76].

## V. CONCLUSIONS

In conclusion, low-temperature interplane heat transport measurements in CeIrIn$_5$ reveal clear anomaly in the configuration with the magnetic field parallel to $a$ axis in the conducting tetragonal plane. The $H-T$ phase diagram of this feature suggests the existence of a new phase inside the superconducting domain, as might be related to the two-component order parameter in this material. Supporting this interpretation, the field dependence of $\kappa_c$ at low temperatures in the $H \parallel c$ configuration shows an activated increase, while that for the $H \parallel a$ configuration is inconsistent with one component models of the superconducting order parameter. This finding shows that the two superconducting states [12] in the phase diagram of Ce(Rh,Ir)In$_5$ are different, with implication that particular type of magnetic order bordering superconductivity is of importance for the superconducting pairing.

## ACKNOWLEDGMENTS

We thank J. Corbin for assistance with the experiment and N. Doiron-Leyraud, F. Laliberte and P. Bourgeois-Hope for critical reading of the manuscript. L.T. acknowledges support from the Canadian Institute for Advanced Research and funding from NSERC, FRQNT, CFI, and the Canada Research Chairs Program. Part of the work was supported by the US Department of Energy (DOE), Office of Basic Energy Sciences, Division of Materials Sciences and Engineering. The research performed at Ames Laboratory, which is operated for the US DOE by Iowa State University under Contract No. DE-AC02-07CH11358. Part of the work was carried out at the Brookhaven National Laboratory, which is operated for the U.S. Department of Energy by Brookhaven Science Associates (No. DE-SC0012704). H.Sh. would like to thank Iran National Science Foundation (INSF) Grant No. 4004583 for supporting this project.